\documentclass[aip,graphicx,amsmath,amssymb,reprint]{revtex4-1}
\usepackage{graphicx}
\usepackage{dcolumn}
\usepackage{bm}
\usepackage{natbib}

\usepackage[utf8]{inputenc}
\usepackage[T1]{fontenc}
\usepackage{mathptmx}

\usepackage{xcolor}

\draft

\begin{document}
\title{Kuramoto model for populations of 
quadratic integrate-and-fire neurons with chemical and electrical coupling}
\author{Pau Clusella}
\affiliation{Department of Experimental and Health Sciences, Universitat Pompeu Fabra, Barcelona Biomedical Research Park, 08003, Barcelona, Spain}
\author{Bastian Pietras}
\affiliation{Institute of Mathematics,  Technical University Berlin,  10623 Berlin,  Germany.}
\affiliation{Bernstein Center for Computational Neuroscience Berlin, 10115 Berlin, Germany.}
\author{Ernest Montbrió}
\affiliation{Neuronal Dynamics Group, Department of Information and Communication Technologies,
Universitat Pompeu Fabra, 08018 Barcelona, Spain.}
\date{\today}

 \begin{abstract}
We derive the Kuramoto model (KM) corresponding to a
population of weakly coupled, nearly identical quadratic integrate-and-fire (QIF) neurons 
with both electrical and chemical coupling. The ratio of chemical to 
electrical coupling determines the phase lag of the characteristic sine coupling function 
of the KM, and critically determines the synchronization properties of the network. 
We apply our results to uncover the presence of chimera states in two coupled populations of 
identical QIF neurons. We find that the presence of both electrical and chemical coupling 
is a necessary condition for chimera states to exist. Finally, we numerically
demonstrate that chimera states gradually disappear as coupling strengths cease to be weak.
\end{abstract}

\pacs{}

\maketitle 

\begin{quotation}
The Kuramoto model (KM) is a minimal mathematical model for 
investigating the emergence of collective oscillations in 
populations of heterogeneous, self-sustained oscillators~\cite{Kur75,Kur84}.
Though the KM model was not originally intended to describe any specific natural system,
an abundant body of work applies it to explore large-scale neuronal oscillations;
see, e.g., Refs.~\onlinecite{GCC+21,LB19,BHD10,CHS+11,VMM14,PDH+15,SKS+15,SLR+15,PSP+16,ARM+16,PPJ18,RGA+19,
CM19,ZZV+20,NPG+20,JEP21,PPM+21,OKD21,PGR21,WDB+21}. 
Yet, it remains unclear how the parameters of the KM relate 
to parameters---such as chemical or electrical synaptic strengths---critical for setting up 
synchronization in biophysically realistic neuronal 
models~\cite{Wan10,Boe17}. 
Here, we unveil a mathematical relation between a popular spiking neuron model, 
the quadratic integrate-and-fire (QIF)~\cite{Erm96,EK86,Izh07}, with 
a well-known variant of the KM~\cite{SSK88,SK86}.   
This provides support in favor of the use of the KM for modeling studies in 
computational neuroscience and introduces the powerful mathematical framework of 
the KM~\cite{PRK01,Str00,PR15i} for the analysis of the dynamics 
of QIF networks. 
\end{quotation}

\section{Introduction}

Large-scale neuronal oscillations emerge due to the synchronous interplay 
of ensembles of neurons. These oscillations are successfully replicated by 
mathematical models of spiking neurons, 
which also allow for a mechanistic understanding of neuronal rhythmogenesis~\cite{Wan10,Boe17}. 
According to these theories, inhibitory synapses play a central role in setting up  
neuronal synchronization either in isolation~\cite{WB96} or due to their 
interplay with excitatory neurons~\cite{WC72}.
Additionally, inhibitory cells are very often coupled electrically, and this 
coupling is usually mediated by so-called gap junctions~\cite{NPR18}. 
Such electrical synapses are well-known to largely favor synchrony. 

Recently, important efforts have been put forward to model the oscillatory dynamics of 
so-called whole-brain networks~\cite{DTB+15,LB19,GCC+21}. 
To facilitate both the analysis and the computational work many studies do not 
use spiking neuron models, but apply the mathematical framework of the 
Kuramoto model (KM);
see, e.g., Refs.~\onlinecite{GCC+21,LB19,BHD10,CHS+11,VMM14,PDH+15,SKS+15,SLR+15,PSP+16,ARM+16,PPJ18,RGA+19,
CM19,ZZV+20,NPG+20,JEP21,PPM+21,OKD21,PGR21,WDB+21}.    
Yet, it remains unclear how to relate the parameters of the KM to bio-physically 
meaningful parameters, such as synaptic strengths.

In this paper, we aim to theoretically substantiate the use of the KM for 
neuronal modeling, by providing a mathematical link between 
the quadratic integrate-and-fire (QIF) model and the KM.
We derive a KM for QIF neurons and subsequently justify its validity in two different ways: 
First, we compare the predictions of the KM with 
those of an exact mean-field model for QIF neurons---often
referred to as firing rate or neural mass model (NMM)~\cite{PDR+19,MP20,MPR15}. 
Second, we use two populations of \emph{identical} Kuramoto 
oscillators to find so-called chimera states~\cite{MKB04,AMS+08,Lai09,MBP16}. 
In a chimera state, one of the two homogeneous populations displays in-phase synchrony, and 
NMMs are useless in this case.
However, the KM for QIF neurons is perfectly suited to describe full synchrony, 
and we exploit this to uncover the existence of chimera states in two-population networks 
of QIF neurons.    

Our derivation of the KM for QIF neurons 
mainly builds on a previous work by Izhikevich~\cite{Izh07}
and also on Refs.~\onlinecite{PM14,MP18}. In Chap. 10 of Ref.~\cite{Izh07}, 
Izhikevich applied perturbation methods to derive 
a simplified model that approximated the dynamics of two 
identical QIF neurons, with \emph{either} chemical or electrical coupling 
\footnote{More recently a similar derivation has been obtained 
for populations of heterogeneous Winfree oscillators with 
sinusoidal infinitesimal Phase Resetting Curves (iPRC)~\cite{PM14}. 
When the oscillators have the iPRC of the QIF neuron, the 
population of Winfree oscillators can be well approximated to 
to a population of QIF neurons with (weak) chemical synapses~\cite{MP18}.}.
Here, we extend the work of Izhikevich and derive a model that 
approximates the dynamics of an \emph{ensemble of heterogeneous} QIF neurons with 
\emph{both} chemical and electrical coupling.
The approximated model turns out to be a well-known version of the 
KM~\cite{SSK88,SK86} and is valid when both heterogeneities 
and coupling strengths are weak. 

This paper is organized as follows: In Section II, we introduce the QIF population model, 
and in Section III, we describe the method to reduce the QIF model to the KM. In Section IV,
we analyze the dynamics of the KM and demonstrate that it correctly describes the 
collective dynamics of populations of nearly identical QIF neurons, with weak electrical and 
chemical synapses. In Section V, we exploit the KM to uncover the presence of 
chimera states in coupled populations of identical QIF neurons. Finally, in Section IV,
we briefly discuss and summarize our results.  

\section{Population of QIF neurons with electrical and chemical 
synapses}

We investigate a population of $N$
quadratic integrate-and-fire (QIF) neurons $i=1,\dots,N$ interacting all-to-all via 
both electrical and chemical synapses~\cite{KE04,Lai15,PDR+19}
\begin{equation}
\tau \dot V_i = V_i^2+\eta_i +\epsilon ~ I_{i,syn}(t), ~~\text{if } V_i>V_p, \text{ then } V_i \leftarrow V_r,
\label{qif}
\end{equation}
where $V_i$ is the membrane potential of neuron $i$,
$\tau$ is the membrane time constant of the neurons, and $\eta_i$ represents an
external current, which varies from cell to cell. 
Due to the quadratic nonlinearity of the 
QIF model, the membrane potential blows up in finite time, and a resetting 
rule is needed: When the neurons reach the peak value $V_p$, they emit a spike and the 
voltage is reset to $V_r$. We assume symmetric 
spike resetting, $V_p=-V_r$ and $V_p \to \infty$, so that the QIF model 
is equivalent to the so-called theta-neuron~\cite{EK86,Erm96}.
In addition, we consider $\eta_i>0$ and hence, in the absence of 
synaptic inputs ($I_{j,syn}=0$), QIF neurons are self-sustained oscillators. 
Finally, synaptic inputs (whose total strength is controlled by the \emph{small} 
parameter $\epsilon\geq 0$) are composed of electrical and chemical synapses,
\begin{equation}
 I_{i,syn}(t) = g (v(t)- V_i) + J \tau  r(t).
\label{Isyn}
\end{equation}
Specifically, electrical synapses (of strength $\epsilon g\geq 0$) diffusively couple 
each neuron with the mean membrane potential 
\begin{equation}
v (t)= \frac{1}{N} \sum_{j=1}^{N} V_j(t).
\label{v}
\end{equation}
Electrical synapses mostly connect inhibitory neurons,
and hence  the chemical synaptic strength, $J$, 
is thought of as a negative parameter thereafter.
Finally, chemical synapses (of strength $\epsilon J$) are mediated by the mean firing rate
\begin{equation}
r (t)= \frac{1}{N} \sum_{j=1}^N\sum_k \delta\left(t- t_j^{(k)}\right),
\label{r}
\end{equation}
where $t_j^{(k)}$ is the time of the $k$th spike of the $j$th neuron and $\delta(t)$ is the Dirac delta function. 

%

\section{Derivation of the Kuramoto model for populations of QIF neurons}

In the following we derive the Kuramoto model corresponding to Eq.~\eqref{qif}. 
The derivation exploits well-known mathematical methods that are reviewed for example 
in Refs.~\onlinecite{HI97,Izh07,PD19}.

We perform the derivation of the KM as follows: First, we obtain the 
phase resetting curve (PRC) of the QIF model. Second, invoking weak coupling, we  
derive the so-called Winfree model corresponding to the QIF model. 
Finally, we assume weak heterogeneity and apply 
the method of averaging to obtain the Kuramoto model corresponding to 
Eq.~\eqref{qif}.  

\subsection{Phase resetting curve (PRC) of a QIF neuron}

We first consider an isolated, regularly spiking QIF neuron; i.e.,~$I_{i,syn}=0$ and $\eta_i > 0$.
The solution of the QIF model immediately after a spike, $V_i(0)=-\infty$, is  
\begin{equation}
V_i(t)=\sqrt{\eta_i} \tan \left( t\sqrt{\eta_i}/\tau -\pi/2\right).
\label{sol}
\end{equation}
The frequency of the oscillations is 
\begin{equation}
\Omega_i=2 \sqrt{\eta_i}/\tau, 
\label{Omega}
\end{equation}
and a phase variable $\theta_i$ can be defined in the interval $[0,2\pi)$ as
\begin{equation}
\theta_i=\Omega_i t=2\arctan( V_i/\sqrt{\eta_i})+\pi.
\label{phase}
\end{equation}
Next, we assume that the neuron is perturbed so that its membrane 
potential instantaneously changes from $V_i$ to $V_i+\delta V$. 
Then, the new phase after the perturbation is 
$\theta_{i,new}=2 \arctan((V_i+\delta V)/\sqrt{\eta_i})+\pi$.

The PRC measures the
phase shift produced by the perturbation, i.e. PRC$=\theta_{i,new}-\theta_i$. 
Hence, for the QIF model, the PRC is~\cite{Izh07}   
\begin{equation}
\text{PRC}(\theta_i,\delta V)=2\arctan(\delta V/\sqrt{\eta_i}-\cot(\theta_i/2))+\pi-\theta_i.
\label{PRC}
\end{equation}
This function depends on both the strength of the perturbation 
and the phase of the neuron at the instant of the perturbation.  
The PRC Eq.~\eqref{PRC} is always positive, indicating that positive/negative 
perturbations only produce positive/negative phase shifts. This 
characterizes the so-called Class 1 neuronal oscillators~\cite{Erm96}.    

\subsection{Weak coupling approximation and the Winfree model}

The PRC Eq.~\eqref{PRC} exactly characterizes the phase response of the 
QIF neuron to a perturbation. Next, we invoke weak coupling, which allows for deriving
a new phase model---called the Winfree model---that
 approximates the network Eq.~\eqref{qif} for $\epsilon \ll 1$.

Weak perturbations produce small changes in the membrane potential, $|\delta V| \ll 1$. 
Then the PRC scales linearly with the strength of the perturbation~\cite{Win67} 
$$\text{PRC}(\theta_i,\delta V) \approx Z(\theta_i) \delta V,$$  
where $Z(\theta_i)$ is called phase sensitivity function or infinitesimal phase resetting curve (iPRC). For the QIF model, the iPRC is 
\begin{equation}
Z(\theta_i)=
\left.\frac{\partial \text{PRC}(\theta_i,\delta V)}{\partial (\delta V)}\right|_{\delta V=0}
=\frac{1-\cos \theta_i}{\sqrt{\eta_i}}.
\label{iPRC}
\end{equation}
When weak perturbations are described by a continuous function $P(t)$
with $|P(t)|\ll 1$, the infinitesimal change in the phase due to the
perturbations is $d \theta = Z(\theta) P(t) dt$.
Accordingly,
assuming weak coupling $\epsilon \ll 1$, 
the population of QIF neurons Eq.~\eqref{qif} is well approximated 
by the Winfree model,
\begin{align}
\dot \theta_i &= \Omega_i + \frac{\epsilon}{\tau} (1-\cos \theta_i)  \sum_{j=1}^N P(\theta_i,\theta_j),
\label{Winfree}
\end{align}
where perturbations to neuron $i$ are due to synaptic inputs from neuron $j$
and can be written in terms of the phase variables as 
\begin{equation}
P(\theta_i,\theta_j)=\frac{g}{N} \left( \cot(\theta_i/2)-\sqrt{\frac{\eta_j}{\eta_i}} \cot(\theta_j/2)\right) +\frac{2J}{N} \sqrt{\frac{\eta_j}{\eta_i}} \delta\left(\theta_j\right).
\label{P}
\end{equation}
Recall that $\theta_j\in [0,2\pi)$ so that 
the Dirac delta function in Eq.~\eqref{P} has argument zero whenever neuron $j$
fires a spike.

\subsection{Weak heterogeneity and the averaging approximation}

The Winfree model can be further simplified using the method of averaging. 
We consider the external currents in Eq.~\eqref{qif} as a common 
current $\bar \eta$ plus a weakly distributed parameter as   
\begin{equation}
\eta_i=\bar \eta + \epsilon \chi_i.
\label{eps}
\end{equation}
In the derivation of the Winfree model, we already assumed weak coupling, $\epsilon \ll 1$. Therefore, the smallness of parameter $\epsilon$ 
implies now the smallness of both coupling terms and the level of heterogeneity. 
This assumption allows for a separation of time scales so that the  
phases $\theta_i$ can be written as   
\begin{equation}
\theta_i=\Phi+\phi_i,
\label{sts} 
\end{equation}
where $\Phi$ describes the fast, free-running oscillation of period 
$$T=\tau \pi/\sqrt{\bar \eta},$$ 
whereas the phases $\phi_i$ describe slow phase drifts produced by 
weak heterogeneities and synaptic inputs.
Substituting Eq.~\eqref{sts} into 
the Winfree model Eqs.~(\ref{Winfree},\ref{P}) and collecting 
terms of order $\epsilon$, 
we find the evolution equation for the slow phases 
\begin{align}
\dot \phi_i =  \frac{ \epsilon\chi_i}{\tau \sqrt{\bar\eta}} + [1-\cos(\Phi+\phi_i)] 
\frac{\epsilon}{\tau N}\sum_{j=1}^N p(\Phi+\phi_i,\Phi+\phi_i+\Delta_{ji} ).
\label{av0}
\end{align}
Here, we defined pairwise phase differences as $\Delta_{ji} = \phi_j-\phi_i$
and a function describing synaptic perturbations as  
\begin{equation}
p( x,y ) = g\left[\cot(x/2)-\cot(y/2)\right]+2J \delta(y).
\nonumber
\end{equation}
To apply the method of averaging to Eq.~\eqref{av0}, 
we consider that in one period of the fast oscillation, $T$, 
the slow phases $\phi_i$ can be assumed constant. Then, Eq.~\eqref{av0} reduces to   
\begin{equation}
\dot \phi_i=\frac{\epsilon}{\tau N} \sum_{j=1}^N \Gamma( \Delta_{ji} ), 
\label{av}
\end{equation}
where the coupling function $\Gamma$ is obtained by averaging 
the r.h.s. of Eq.~\eqref{av0} over one period $T$. 
This involves the evaluation of four integrals that can be 
explicitly computed and yields the phase interaction function  
\begin{equation}
\Gamma(\Delta_{ji})=  \frac{ \chi_i}{\sqrt{\bar\eta}} + 
g \sin \Delta_{ji}+  \frac{J}{\pi} 
( 1-\cos \Delta_{ji}). 
\label{gamma}
\end{equation}
%
\subsection{Kuramoto model for populations of QIF neurons}

Substituting Eq.~\eqref{gamma} into Eq.~\eqref{av} and expressing the result in terms of the 
original phases Eq.~\eqref{sts}, we find the Kuramoto model  
\begin{equation}
\dot \theta_i= \omega_i+ 
\frac{\epsilon}{\tau N} \sum_{j=1}^N \left[g \sin ( \theta_j-\theta_i)- \frac{J}{\pi} 
\cos( \theta_j-\theta_i) \right] +\frac{\epsilon }{\tau}  \frac{J}{\pi}
\label{KM}
\end{equation}
with natural frequencies
\begin{equation}
\omega_i =  \frac{2 \sqrt{\bar\eta}}{\tau} +\epsilon \frac{\chi_i}{\tau\sqrt{\bar\eta}}.
\label{om}
\end{equation}
In the absence of electrical synapses $g=0$, Eqs.~\eqref{KM} essentially 
\footnote{For $g=0$ the natural frequencies  
in Refs.~\cite{PM14,MP18} differ from Eq.~\eqref{om}. 
The reason for this discrepancy is that Refs.~\cite{PM14,MP18} 
consider the Winfree model with distributed natural frequencies $\Omega_i$,
while here we study the QIF model with distributed currents $\eta_i$.}
reduces to the Kuramoto model with chemical synapses derived in Refs.~\onlinecite{PM14} and~\onlinecite{MP18}. 
The KM for QIF neurons Eq.~\eqref{KM} generalizes the results
in Refs.~\onlinecite{PM14} and~\onlinecite{MP18} to networks with both electrical and 
chemical coupling, and it is our main result.

Eq.~\eqref{om} is the linear approximation of Eq.~\eqref{Omega} for 
weak heterogeneity---see also Eq.~\eqref{eps}. 
The last term of Eq.~\eqref{KM}  
describes the deviation of the natural frequencies due to synaptic coupling, 
which exclusively depends on chemical coupling. 
Excitatory coupling ($J>0$) speeds up the frequencies of the oscillators, 
and inhibition ($J<0$) slows them down. 
These frequency shifts do not qualitatively affect the collective dynamics of 
Eq.~\eqref{KM}, but they may become relevant 
if the oscillators are not all-to-all coupled~\cite{SSK88,BT05} 
or in the case of 
interacting excitatory and inhibitory populations~\cite{MP18}.

\begin{figure}[t]
\includegraphics[width=0.35\textwidth]{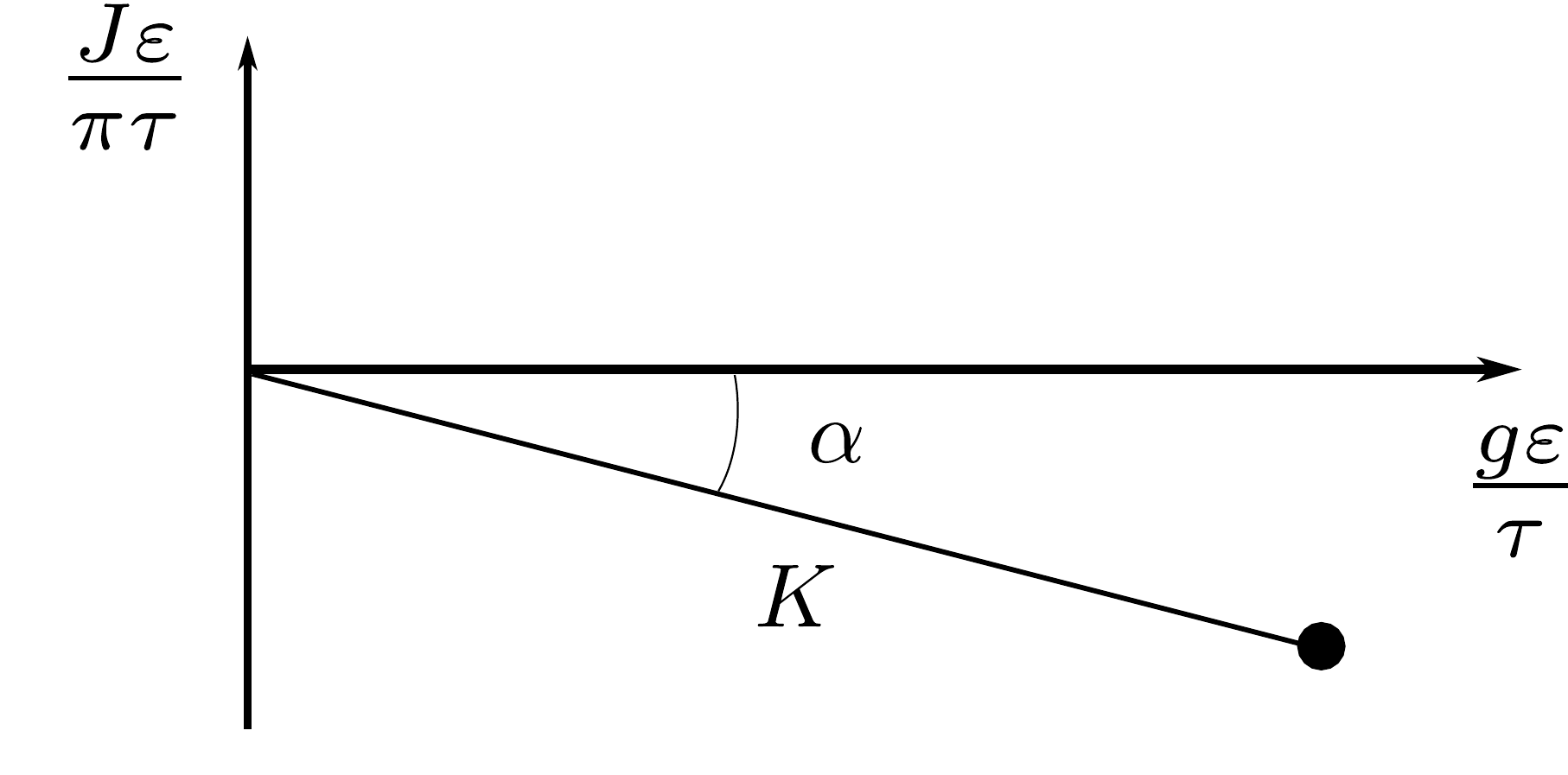} 
\caption{Geometric relation---determined by Eqs.~(\ref{K},\ref{alpha})---between the coupling parameters of the QIF model Eqs.~\eqref{qif} 
and the Kuramoto model Eqs.~\eqref{KMp}. }
\label{Fig1}
\end{figure}

Alternatively, Eq.~\eqref{KM} can be cast in the more transparent  
form~\cite{SSK88,SK86}
\begin{equation}
\dot \theta_i =\omega_i+\frac{K}{N}\sum_{j=1}^N 
\left[\sin(\theta_j-\theta_i -\alpha)+\sin\alpha \right],
\label{KMp}
\end{equation}
with the coupling constant 
\begin{equation}
K=\frac{\epsilon}{\tau}\sqrt{(J/\pi)^2+g^2},
\label{K}
\end{equation}
and the phase lag parameter 
\begin{equation}
\alpha=\arctan\left(\frac{J/\pi}{g}\right).
\label{alpha}
\end{equation}
The coupling parameters $K$ and $\alpha$ satisfy a simple geometric relation with the 
coupling parameters of the QIF model Eq.~\eqref{qif}, illustrated 
in Fig.~\ref{Fig1}.  
Given a particular choice of the QIF coupling parameters, 
Fig.~\ref{Fig1} shows that 
electrical coupling and chemical coupling (divided by a factor $\pi$) 
contribute equally to the overall coupling strength $K$ of the KM, 
and that $K$ is insensitive to the sign of the chemical coupling---in Fig.~\ref{Fig1} we consider an inhibitory network; i.e., $J<0$.

To lighten the notation, we consider $\epsilon=1$ thereafter. Hence, for the 
KM Eq.~\eqref{KMp} to be a good approximation of Eq.~\eqref{qif},  
in the following, the synaptic weights $J$ and $g$ need to be regarded as small quantities.

\section{Analysis of the Kuramoto model for Quadratic Integrate-and-fire neurons}

Using Fig.~\ref{Fig1}---or, equivalently, Eq.~\eqref{alpha}---we may infer how chemical and electrical synapses contribute to synchronization, 
using well-known results for the KM.
For example, the phase constant $\alpha$ 
critically determines the synchronization behavior of Eq.~\eqref{KMp}~\cite{SK86}.
In the absence of electrical coupling, $g= 0$, we find $\alpha = +\pi/2$
for excitatory coupling and $\alpha=-\pi/2$ for inhibitory coupling. This 
indicates that collective synchronization is unreachable---consistent 
with the well-known fact that instantaneous chemical coupling
is unable to synchronize type 1 neuronal oscillators~\cite{Erm96,HMM95,DRM17}.
In contrast, in the absence of chemical coupling, one finds $\alpha=0$, 
and Eq.~\eqref{KMp} reduces to the standard KM, in which 
collective synchronization is   
achieved at a critical degree of heterogeneity $\Delta=\Delta_c(K)$ 
that depends on the coupling strength~\cite{Kur84}.
Between these two extreme cases, that is in networks 
with both electrical and chemical synapses, 
we find the phase lag parameter $|\alpha| \in (0,\pi/2)$,
and synchronization generally depends on both $\alpha$ and the 
overall shape of the distribution of natural frequencies~\cite{SK86}. 

To validate the KM for QIF neurons, in Section IV A,
 we obtain the mean-field model corresponding to Eqs.~\eqref{KMp} and compare its predictions with those of the mean-field model derived
in Ref.\cite{PDR+19}, which describes the dynamics of the QIF network
Eq.~\eqref{qif} exactly.

\subsection{Mean-field model}

\begin{figure}[b]
\includegraphics[width=85mm]{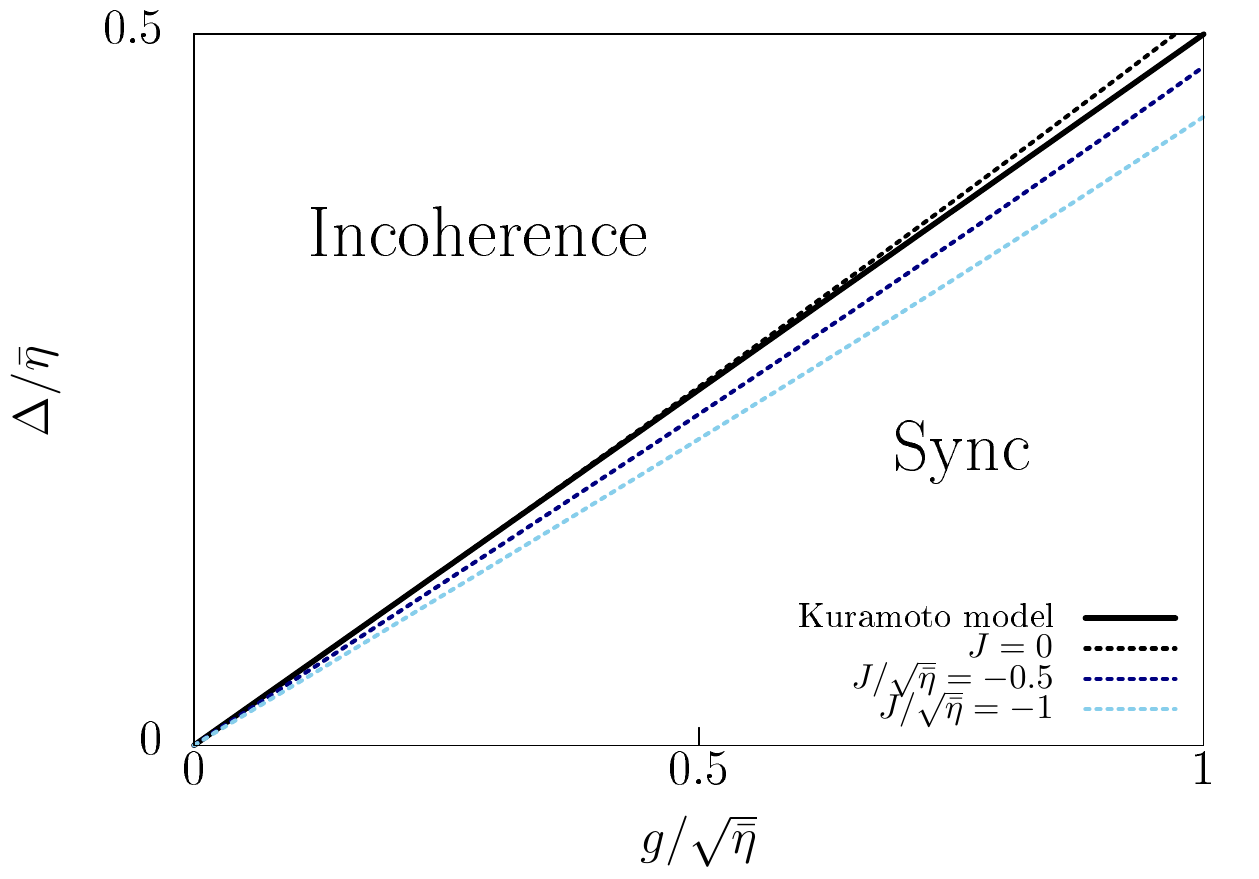} 
\caption{Synchronization boundary of the QIF network Eq.~\eqref{qif} 
with the Lorentzian distribution of currents for various 
values of the (scaled) inhibitory coupling strength, $J/\sqrt{\bar \eta}$. 
The solid line corresponds to 
the approximated critical width Eq.~\eqref{DeltaC}, which is independent of $J$. 
Dashed lines correspond to the exact synchronization boundaries, obtained 
using Eq.~(7) in Ref.~\cite{PDR+19}.}
\label{Fig2}
\end{figure}

In the thermodynamic limit ($N \to \infty$), the dynamics of
Eqs.~\eqref{KMp} are greatly simplified assuming    
$\chi_i$ in Eq.~\eqref{om} to be Lorentzian-distributed 
$$G(\chi)= \frac{\Delta/\pi}{\chi^2+\Delta^2},$$ 
where $\Delta$ is the half-width of the distribution.
Then, using the so-called Ott-Antonsen (OA) ansatz~\cite{OA08},
the KM Eqs.~\eqref{KMp} can be exactly reduced to a
mean-field model consisting of two differential equations
for the complex Kuramoto order parameter,
\begin{equation}
Z=R e^{i \psi}= \frac{1}{N}\sum_{j=1}^N e^{i \theta_j},
\label{kop0}
\end{equation} 
in the limit $N\to\infty$. The mathematical approach to obtain the mean-field equations 
corresponding to Eqs.~\eqref{KMp} is a standard procedure. 
Here, we skip the mathematical details and refer the reader to, for example   
Ref.~\cite{MP11}, where the 
mean-field model corresponding to Eq.~\eqref{KMp} was derived in detail. 
Accordingly, using Eqs.~(\ref{om},\ref{K},\ref{alpha}), 
we obtain the mean-field equations 
\begin{subequations}
\label{OA0}
\begin{eqnarray}
\dot R &=& \frac{R}{2\tau} \left(- \frac{2\Delta}{\sqrt{\bar\eta}}+g(1-R^2) \right),
\label{R}\\
\dot \psi &=& \frac{ 2 \sqrt{\bar\eta}}{\tau}+ \frac{J}{2\pi\tau } (1-R^2) ,
\label{psi}
\end{eqnarray}
\end{subequations}
which approximate the dynamics of the QIF model Eqs.~\eqref{qif} 
for small $g$ and $J$.
The radial equation Eq.~\eqref{R} shows that the incoherent state ($R=0$) 
is a stable fixed point above the critical width, 
\begin{equation}
\Delta_c =  g \sqrt{\bar \eta}/2 ,
\label{DeltaC}
\end{equation}
which is independent of chemical coupling, $J$.
At $\Delta=\Delta_c$ a stable nontrivial solution---corresponding to a partially 
synchronized state---bifurcates from incoherence with 
\begin{equation}
R= \sqrt{(\Delta_c-\Delta)/\Delta_c},
\nonumber
\end{equation}
and frequency 
\begin{equation}
\Omega= \frac{2\sqrt{\bar \eta}}{\tau}+ \frac{\Delta}{\tau \pi\sqrt{\bar \eta}}\frac{J }{g }.
\label{OmC}
\end{equation}
The solid line in Fig.~\ref{Fig2} corresponds to the critical boundary Eq.~\eqref{DeltaC},
while dashed lines correspond to the exact synchronization 
boundaries of the QIF network for various degrees of inhibitory coupling---see Eq.~(7) in Ref.~\onlinecite{PDR+19}.
Note that for weak electrical coupling and/or weak heterogeneity, 
all boundaries approach Eq.~\eqref{DeltaC}. Furthermore, for weak heterogeneity, 
the frequency of the synchronized cluster Eq.~\eqref{OmC} agrees
with Eq.~(8) in Ref.~\cite{PDR+19}, which describes the frequency of 
the oscillations near their onset.  

In sum, these results confirm  
the validity of the Kuramoto model Eq.~\eqref{KMp} as an approximation of a
population of heterogeneous QIF neurons with electrical and chemical 
coupling, Eq.~\eqref{qif}.

\section{Chimera states in coupled homogeneous populations of QIF neurons}

To further illustrate the appropriateness of the KM 
to investigate the dynamics of QIF networks, 
we investigate the presence of chimera states 
in populations of QIF neurons. Our motivation is threefold:  

\begin{enumerate}
\item Chimera states were originally uncovered 
in a nonlocally coupled network of identical Kuramoto oscillators~\cite{KB02}.
Given that the phase dynamics Eqs.~\eqref{KMp} are an approximation of Eqs.~\eqref{qif}
valid for weak heterogeneity and weak coupling, 
we expect QIF networks to display 
similar chimera states, at least for weak coupling. 

\item Several papers have been devoted to investigate 
chimera states in networks of spiking neurons;
see, e.g., Refs.~\onlinecite{Sak06,OPT10,OOH+13,VHO+14,HKV+14,
PA15,MBG+19,Lai19,HKZ+16,MBG+19,GBS+20,LJH21}.
Some of them provide
numerical evidence that the presence of both chemical and electrical synapses favors the emergence of chimera states~\cite{HKZ+16,MBG+19}.
Yet, the relation between chimera states in spiking neuron networks 
with the original chimera states uncovered in the KM~\cite{KB02,MKB04,AMS+08,Lai09} 
is lacking~\cite{BGL+20,Ome18,Hau21}. 

\item Recently, exact mean-field models for large populations of QIF 
neurons (often called neural mass models, NMMs) 
with electrical and chemical synapses have been put forward
\cite{Lai15,PDR+19,MP20,BRN+20}. However, such NMMs 
have an important limitation when neurons are---as in a chimera state---identical and fully synchronized since both the 
mean membrane potential and the  
mean firing rate diverge at the instant of collective firing
\footnote{
NMMs for QIF neurons are derived after adopting the thermodynamic limit, $N \to \infty$, and 
under the assumption $V_r \to -\infty$, and $V_p \to \infty$, see \cite{MP20,MPR15}.
If neurons are fully synchronized, ($V_i(t)=V_j(t), ~\forall i,j$)
the population of QIF neurons behaves as a single neuron. 
Therefore, in QIF-NMM, the mean voltage Eq.~\eqref{v}  
diverges when all neurons fire a spike---in Refs.~\cite{Lai15,Erm06} the mean-field variable $v$ is approximated to avoid this divergence, see also \cite{PDR+19}. }.
This divergence is avoided using the averaging approximation, and hence   
the KM for QIF neurons, Eq.~\eqref{KM}, becomes singularly suited to study 
collective behavior where neurons are fully synchronized. %
\end{enumerate}

Chimera states were originally uncovered 
in a ring of identical Kuramoto oscillators with nonlocal coupling
when $\alpha \lesssim \pi/2$~\cite{KB02}. 
Shortly after their discovery, chimera states were also found in a simpler setup, 
consisting of two populations of identical 
Kuramoto oscillators~\cite{MKB04,AMS+08}. 
Here, to investigate chimera states in networks of QIF neurons, 
we adopt the two-population setup of Refs.~\onlinecite{MKB04,AMS+08}.

Specifically, we analyze the dynamics of two identical populations 
(labeled $\sigma \in \{1,2\}$) of $n=N/2$ identical QIF neurons, 
interacting all-to-all via both chemical and electrical synapses 
\begin{equation}
\tau \dot V_i^{\sigma} = \left( V_i^{\sigma} \right)^2+\bar \eta +I_{i,syn,s}^{\sigma}+
I_{i,syn,c}^{\sigma}, 
\label{qif2}
\end{equation}
with the resetting rule of Eq.~\eqref{qif}. Synaptic inputs have a contribution 
$I_{i,syn,s}^{\sigma}$ due to 
self-interactions within each population $\sigma$
and another contribution  $I_{i,syn,c}^{\sigma}$ due to cross-interactions 
of population $\sigma=\{1,2\}$ with population $\sigma'=\{2,1\}$,
\begin{eqnarray*}
I_{i,syn,s}^{\sigma} &=& g_s (v^{\sigma}- V_i^{\sigma}) + J_s \tau  r^{\sigma},\\
I_{i,syn,c}^{\sigma} &=& g_c (v^{\sigma'}- V_i^{\sigma}) + J_c \tau  r^{\sigma'}.
\end{eqnarray*}
Here, $v^{\sigma}$ and $r^{\sigma}$ are the mean 
membrane voltage and mean firing rate of population $\sigma$, respectively. 
Using Eqs.~(\ref{KMp},\ref{K},\ref{alpha}), it is straightforward to write the 
KM corresponding to Eqs.~\eqref{qif2} as 
\begin{eqnarray}
\dot \theta_i^{\sigma}= \omega
		&+& \frac{K_s}{n}\sum_{j=1}^n
\left[\sin\left(\theta_j^{\sigma}-\theta_i^{\sigma} -\alpha_s \right) 
+\sin \alpha_s    \right]\nonumber\\
                &+& \frac{K_c}{n}\sum_{j=1}^n 
\left[ \sin\left(\theta_j^{\sigma'}-\theta_i^{\sigma} -\alpha_c \right)
+\sin \alpha_c    \right],
\label{KM2}
\end{eqnarray}
with $\omega = 2 \sqrt{\bar\eta}/\tau$ and
\begin{eqnarray}
K_{s,c}&=&\frac{1}{\tau}\sqrt{(J_{s,c}/\pi)^2+g_{s,c}^2}, \label{Ksc}\\
\alpha_{s,c}&=&\arctan\left(\frac{J_{s,c}/\pi}{g_{s,c}}\right).  \label{alphasc} 
\end{eqnarray}

The KM Eqs.~\eqref{KM2} is slightly more general than the model originally investigated 
in Refs.~\cite{MKB04,AMS+08,Lai09}---which considered $\alpha_c=\alpha_s$.  
In the QIF network, this equality of the phase lag parameters implies that the ratios of chemical to electrical coupling 
\begin{equation}
\rho_{s} =\frac{J_{s}/\pi}{g_{s}},\quad \rho_{c} =\frac{J_{c}/\pi}{g_{c}},
\label{rho}
\end{equation}
are identical, $\rho_s=\rho_c.$ Recent work has also 
considered the dynamics of chimera states in populations of Kuramoto oscillators 
with distributed phase lags~\cite{MBP16,CRK16}.
Specifically, Martens et al.~\cite{MBP16} investigated chimera 
states in the two-population model Eqs.~\eqref{KM2}.

\subsection{Mean-field model}

As we discussed previously, in the thermodynamic limit ($n=N/2 \to \infty$) 
the KM can be exactly reduced to a low-dimensional mean-field model
using the OA ansatz. In the case of the 
homogeneous, 
two-population Kuramoto model Eqs.~\eqref{KM2},
the dynamics reduces to six ordinary differential equations using the 
Watanabe-Strogatz ansatz~\cite{PR08,WS94}.
Assuming a particular set of initial conditions 
for the phases, the system further reduces to four differential equations
and it is described by the OA ansatz~\cite{PR08,AMS+08}. 
Such mean-field equations describe the evolution of the complex 
Kuramoto order parameters of the two populations,
\begin{equation}
Z_\sigma=R_{\sigma}e^{i \psi_\sigma}= \frac{1}{n}\sum_{j=1}^n e^{i \theta^\sigma_j}.
\label{kops}
\end{equation} 
Using the mean-field analysis in Refs.~\onlinecite{KNA+10,MBP16} and  
Eqs.~(\ref{Ksc},\ref{alphasc},\ref{rho}), the mean-field equations for the complex Kuramoto 
order parameters can be further reduced (by virtue of the rotational symmetry of the KM)
to the three dimensional system 
\begin{subequations}
\label{OA}
\begin{eqnarray}
\frac{d R_1}{d\tilde t}&=& \frac{1-R_1^2}{2}\left[ R_1 + \frac{g_c}{g_s} R_2  \cos \Psi 
	- \rho_s\frac{J_c}{J_s} R_2 \sin \Psi \right], \label{R1}\\
\frac{d R_2}{d\tilde t} &=& \frac{1-R_2^2}{2}\left[ R_2 + \frac{g_c}{g_s}  R_1 \cos \Psi 
	+ \rho_s\frac{J_c}{J_s}  R_1   \sin \Psi \right],\label{R2} \\
\frac{d\Psi}{d\tilde t}&=& \rho_s \frac{ R_1^2-R_2^2 }{2R_1R_2} 
		\left(\frac{J_c}{J_s} \cos \Psi -R_1R_2 \right)-\nonumber \\
		&&\frac{g_c}{g_s}\frac{R_1^2+R_2^2+2 R_1^2 R_2^2}{2R_1R_2}\sin \Psi,
\label{phi}
\end{eqnarray}
\end{subequations}
where the phase difference between the complex 
order parameters Eq.~\eqref{kops} is defined as $\Psi=\psi_1-\psi_2$. 
In addition, we have rescaled time as $\tilde t = g_s t/\tau$ so that the 
dynamics of Eqs.~\eqref{KM2} depends only on three combinations of parameters: 
the ratios of cross to self couplings $g_c/g_s$ and $J_c/J_s$
and the ratio of chemical to electrical coupling $\rho_s$ ---see Eq.~\eqref{rho}. 
In contrast, the original QIF model Eq.~\eqref{qif2} can, after appropriate rescaling,
only be reduced to involve at least four parameters.

\subsection{Phase diagram of the mean-field model}

Chimera states in two-population Kuramoto networks 
correspond to symmetry-broken 
states where one of the populations is fully synchronized (i.e. $R_\sigma=1$),
while the other remains only partially synchronized ($R_{\sigma'}<1$). 
In addition, chimera states in two-population networks of identical Kuramoto 
oscillators coexist with the fully synchronized state,
$R_1=R_2=1$~\cite{MKB04,AMS+08,Lai09}.

\begin{figure}[t]
\includegraphics[width=0.5\textwidth]{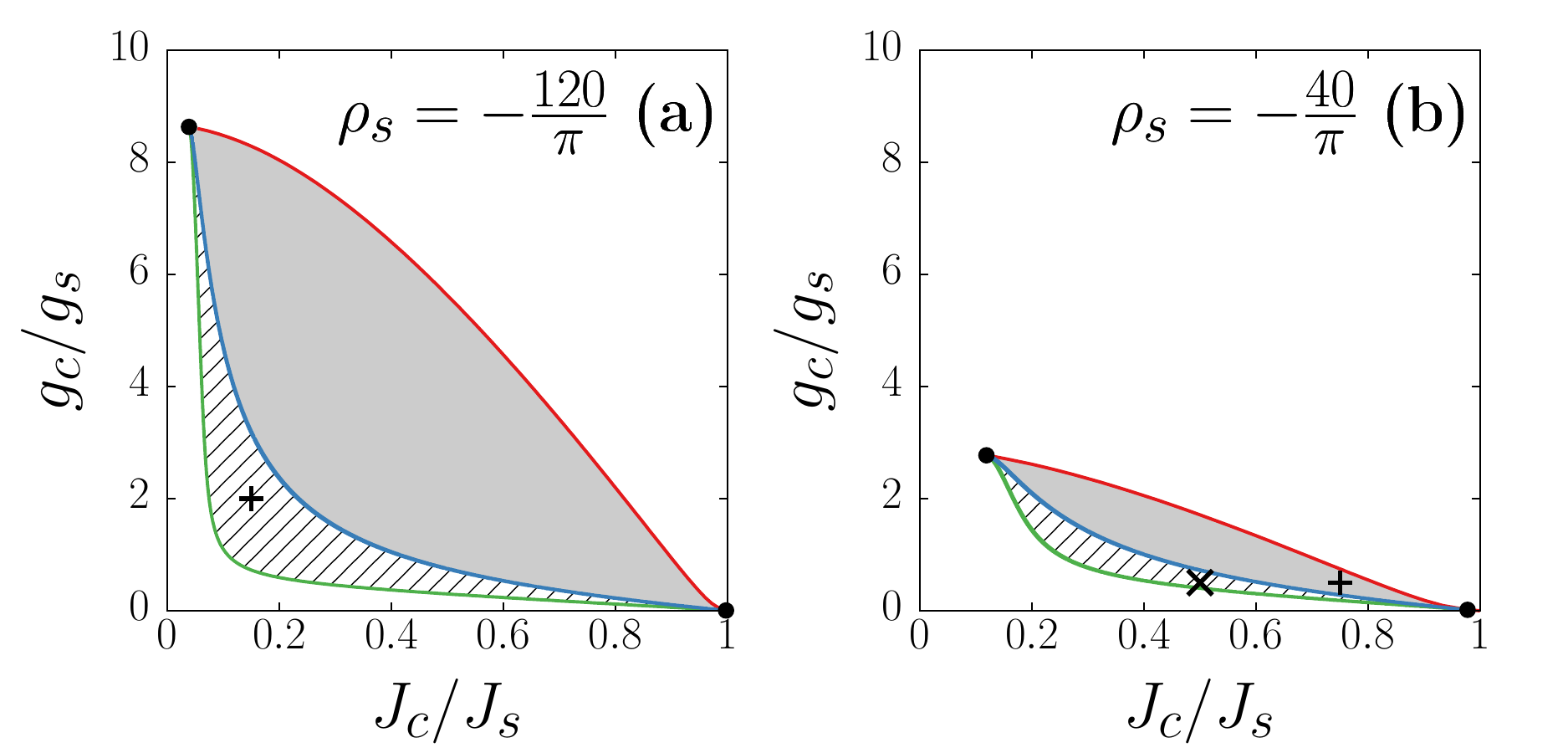} 
\caption{Phase diagrams of the mean field model~Eqs.~\eqref{OA}, for two 
values of the ratio $\rho_s$; see Eqs.~\eqref{rho}. Shaded and hatched regions
correspond to regions of steady and unsteady stable chimera states, respectively.
Red lines: Saddle-node (SN) bifurcations. Blue lines: Hopf bifurcations.
Green lines: Homoclinic bifurcations. Filled circles: Takens-Bogdanov 
points. 
In panel (a), the $+$ symbol corresponds to the coordinates used
for the numerical simulations of the QIF network depicted in Fig.~\ref{Fig4}(f,g,h):
$J_c/J_s=0.15$, $g_c/g_s=2.0$.
In panel (b), symbols correspond to the coordinates used 
for the numerical simulations of the QIF network depicted in 
Fig.~\ref{Fig4}(b,c,d) and Fig.~\ref{Fig5}:  
$J_c/J_s=0.75$, $g_c/g_s=0.5$ ($+$ symbol)
and in Fig.~\ref{Fig6}:  $J_c/J_s=0.5$, $g_c/g_s=0.5$ ($\times$ symbol).  }
\label{Fig3}
\end{figure}

\begin{figure*}
\includegraphics[width=1.\textwidth]{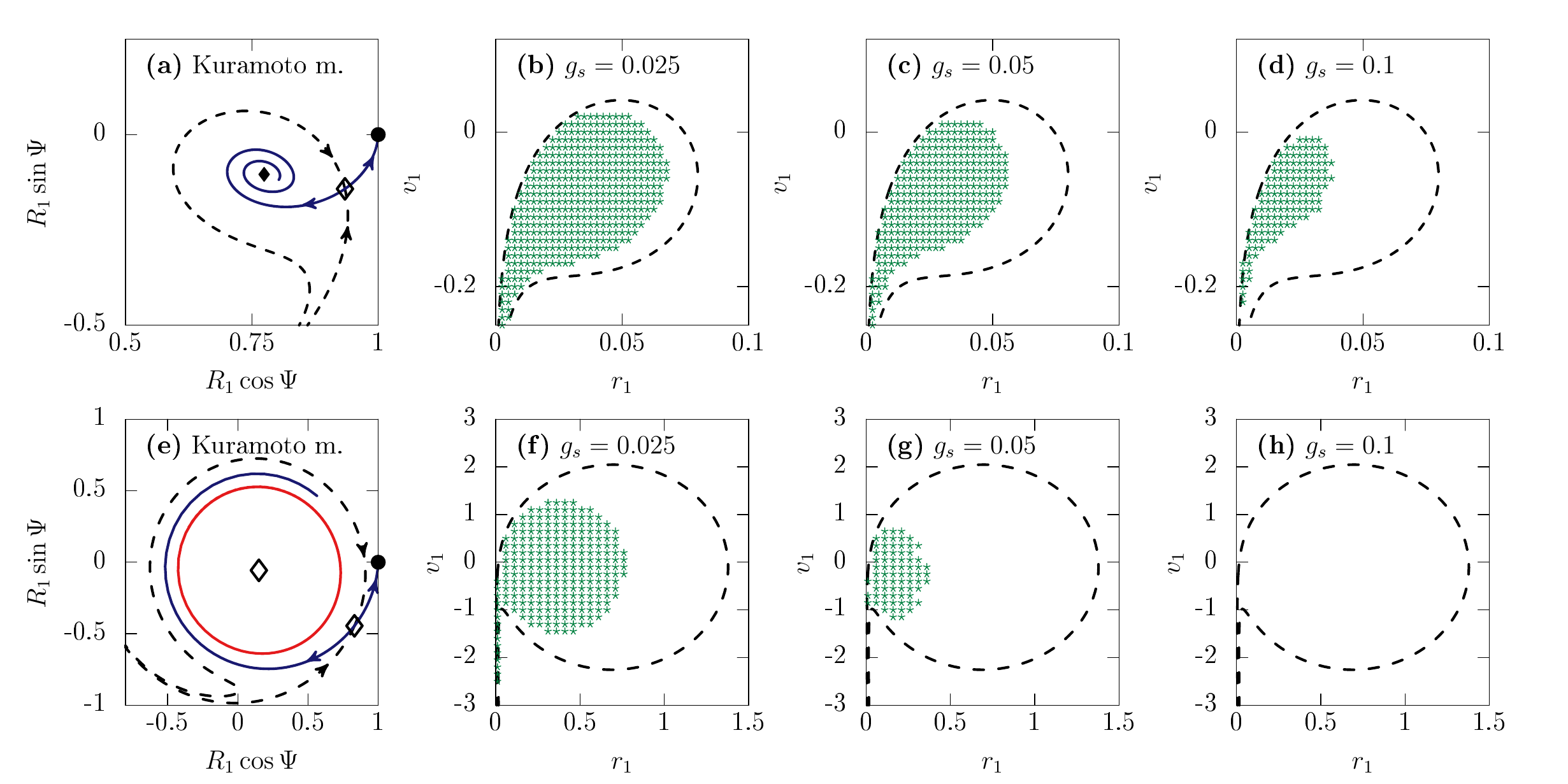}
\caption{Basins of attraction of chimera states in the two-population KM (dashed lines) 
and in the two-population QIF model (green-dotted regions). 
Panels (a-d): Phase portraits of the (a) KM and (b-d) QIF model, with 
$J_c/J_s=0.75$, $g_c/g_s=0.5, \rho_s=-40/\pi$, see $\times$ symbol in Fig.~\ref{Fig3}(b). 
Panels (e-h): Phase portraits of the (e) KM and (f-h) QIF model, with 
$J_c/J_s=0.15$, $g_c/g_s=2, \rho_s=-120/\pi$, see  $+$ symbol in Fig.~\ref{Fig3}(a).
Symbols and lines in panels (a,e): Solid/open diamond: 
Stable/unstable chimera states, respectively; 
solid dot: In-phase synchronized state. 
Solid blue line: Unstable manifold of the saddle point. 
Dashed black lines: Stable manifold of the saddle point, 
basin of attraction of stable chimeras. 
The basins of attraction of chimeras 
have been transformed from the $(R_1 \cos \Psi,R_1\sin \Psi)$ coordinates (panels a,e) 
to the $(r_1,v_1)$ coordinates (panels b-d and f-h) using Eq.~\eqref{conformal}.
Green dots corresponds to initial values leading to 
chimera states after $t=2500$ time units in numerical simulations of 
two populations of $n=200$ QIF neurons.    
Parameters: $\tau=1$ and $\eta=1$.
Simulations using the Euler scheme with time step: 
$dt=10^{-4}$ and symmetric resetting: $V_p=-V_r=1000$.}
\label{Fig4}
\end{figure*}

To obtain the phase diagrams depicted in Fig.~\ref{Fig3},
we set $R_2=1$ in Eqs.~\eqref{OA} and numerically continued~\cite{AUTO} chimera states 
using initial conditions in their basin of attraction---see Fig.~\ref{Fig4}(a) 
and Ref.~\onlinecite{AMS+08}. 
The diagrams show the regions where steady (shaded) and 
unsteady (hatched) chimera states are stable
for two different values of the ratio $\rho_s$; see Eqs.~\eqref{rho}. 
These regions lie between a saddle-node (red) and a Homoclinic (green) bifurcation lines, 
which---together with a Hopf (blue) bifurcation line
separating steady and unsteady chimera states---meet at two  
Takens-Bogdanov (TB) points. 
For decreasing $|\rho_s|$, the region of chimera states  
shrinks and eventually disappears when the two TB points collide. 

The phase diagrams in Fig.~\ref{Fig3} are qualitatively 
identical to that of Fig.~4a in Ref.~\onlinecite{MBP16}, but here the 
regions of chimeras are represented 
in the parameter space of the QIF model
\footnote{Figures~4(b-f) in Ref.~\cite{MBP16} show a bifurcation scenario with 
a transcritical bifurcation that is not observed in Fig.~\ref{Fig3}. 
This bifurcation occurs for $\alpha_s>\pi/2$, and 
these values of the phase lag parameter are unreachable in the QIF model, 
where $\alpha_s\in [-\pi/2,0)$ for inhibitory coupling and 
$\alpha_s \in (0,\pi/2]$ for excitatory coupling), see Eq.~\eqref{alphasc}.}. 
This allows us to determine three   
necessary conditions for the existence of chimera states in two-population 
networks of QIF neurons:
\begin{enumerate}

\item Chimera states only exist in the 
presence of \emph{both} chemical and electrical coupling. 
\item Self-chemical coupling needs to be much larger 
than self-electrical coupling, $|J_s| \gg g_s$ ---or, equivalently, $|\rho_s| \gg 1$.
Using  Eq.~\eqref{alphasc}, this implies that $\alpha_s$ is close to 
$\pm \pi/2$ in correspondence with 
previous work~\cite{MBP16,MKB04,AMS+08}.
\item The modulus of self-chemical coupling needs to be larger than that of 
cross-chemical coupling, $|J_s|>|J_c|$. 
\end{enumerate}

These three conditions are not sufficient conditions to have chimera states in 
Eqs.~\eqref{qif2} though. Indeed, Eq.~\eqref{KM2} and the corresponding 
mean-field Eqs.~\eqref{OA} are an approximation of the QIF 
Eqs.~\eqref{qif2} for weak coupling, but  
it remains to be seen whether chimera states persist in QIF networks 
when coupling strengths become stronger. We numerically explore this 
issue in the next sections. 


\subsection{Chimera states in populations of QIF neurons}

In the following, we numerically investigate the presence of chimera states in the 
spiking neuron network model of QIF neurons
Eqs.~\eqref{qif2}. First, we confirm that, for weak coupling,
chimeras are present in QIF networks and they 
exist in the parameter range predicted by the phase diagrams of the KM,
Fig.~\ref{Fig3}. However, then we show that 
the basin of attraction of chimera states shrinks 
as synaptic coupling strengths become stronger.

\subsubsection{Dynamics of chimera states}

Using the OA ansatz, the dynamics of the two-population model 
Eqs.~\eqref{KM2} with $N\to \infty$ can be exactly reduced to the
three-dimensional system Eqs.~\eqref{OA}. 
In a chimera state we may set $R_2=1$ so that  
Eqs.~\eqref{OA} further reduce to a planar system with variables $R_1$ and $\Psi$. 
Figures~\ref{Fig4}(a,e) show the basins of attraction (dashed lines) of (a) a steady 
chimera state (solid diamond symbol) and of
(e) an unsteady chimera state (red limit cycle), 
with parameters corresponding to $+$ symbols in Fig.~\ref{Fig3}.
As mentioned previously, chimera states coexist with the stable fully synchronized 
solution (solid dot symbols), $Z_1=R_1=1,~\Psi=0$. The basin of attraction of the chimera 
state is defined by the stable manifold of a saddle point (open diamond symbols)
~\cite{AMS+08}.

To set initial conditions leading to the chimera state of Fig.~\ref{Fig4}(a) 
in the network of QIF neurons, 
we considered the initial condition $Z_2=1$ and $Z_1=R_1 e^{i\Psi}$
---with $R_1$ and $\Psi$ such that the system is in 
the basin of attraction of the steady chimera state. 
Then we used the conformal map~\cite{MPR15}
\begin{equation}
\label{conformal}
\pi r_\sigma - i v_\sigma =\frac{1-Z_\sigma}{1+Z_\sigma},
\end{equation}
to transform the mean-field coordinates $Z_2$ and $Z_1$ into the mean firing rate $r$
and the mean membrane potential $v$ of the populations of QIF neurons---for population 2 we find $\pi r_2 + i v_2=0$. Then, 
we initialized the membrane voltages of the populations according to the formula 
$V_i(0)=v+(\pi \tau r) \tan[\pi/2 (2i-n-1)/(n+1)],$ for $i=1,\dots,n$~ 
\footnote{
For $n\to \infty$, this corresponds to a Lorentzian distribution of voltages 
$\rho(V)=\pi \tau r /[ (V-v)^2+ (\pi \tau r)^2]$; see Ref.~\onlinecite{MPR15},
and to uniformly distributed
constants of motion in the Watanabe-Strogatz theory~\cite{PR08,Lai18,BGL+20}
}.

\begin{figure}[t]
\centerline{\includegraphics[width=85mm,clip=true]{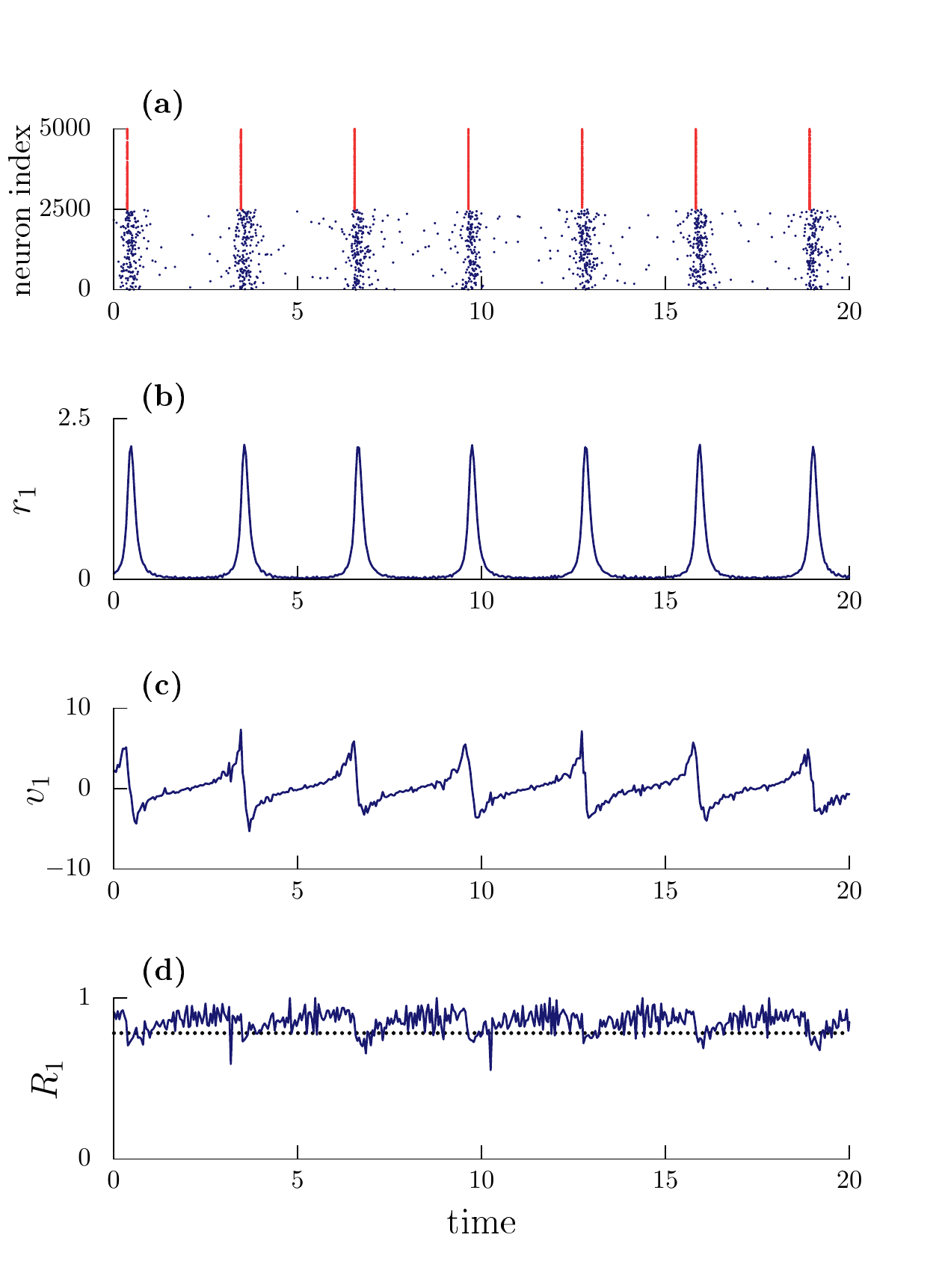}}
\caption{Periodic chimera state in a 
two-population network of $N=5000$ identical
inhibitory QIF neurons ($n=2500$ neurons in each population).
(a) Raster plot of 500 randomly chosen neurons. Neurons in population 1 (blue) are
partially synchronized, and neurons on population 2 are in-phase synchronized (red).
(b) Time series of the mean firing rate $r_1$ of population 1,
computed averaging the firing rate at each time step 
in time windows of $\delta t=0.05$. 
(c) Time series of the mean membrane potential $v_1$
of population 1. 
(d) Time series of the Kuramoto order parameter of population 1, $R_1$,
obtained from the mean-field quantities $r_1,v_1$ using the conformal map 
Eq.~\eqref{conformal} (blue lines) 
and from direct integration of Eqs.~(\ref{OA}) (black dots).
Parameters as in Fig.~\ref{Fig4}d (see also $+$ symbol in Fig.\ref{Fig3}b): 
$J_c/J_s=0.75$, $g_c/g_s=0.5$, $\rho_s=-40/\pi$, and $g_s=0.1$.
Numerical simulations performed using the Euler scheme with time step: 
$dt=10^{-4}$ and symmetric resetting: $V_p=-V_r=1000$.
}
\label{Fig5}
\end{figure}

In Fig.~\ref{Fig5}, we show the results of a numerical simulation of the QIF network
($g_s=0.1$).
The raster plot in Fig.~\ref{Fig5}(a) clearly shows the signature of a chimera state: 
Neurons in population 1 (blue) are only partially synchronized, while 
neurons in population 2 remain fully synchronized.
The time evolution of the firing rate $r_1$ and the mean-membrane 
potential $v_1$ for the incoherent group are displayed in 
Figs.~\ref{Fig5}(b) and (c), respectively. 
These collective variables indicate a periodic evolution of the incoherent population,
with fluctuations caused by the  
finite resetting of the QIF neurons and by finite-size effects.  
Finally, Fig.~\ref{Fig5}(d) shows the Kuramoto order parameter $R_1$ (blue) 
obtained using the time series $r_1(t)$ and $v_1(t)$ and 
the conformal map Eq.~\eqref{conformal}. 
In contrast with the steady chimera state in the mean-field Eqs.\eqref{OA}
(black dotted line),
the chimera state in the network of QIF neurons is not stationary but oscillates 
periodically in time. The same unsteady chimeras arise in 
two-population networks of Winfree oscillators~\cite{PM14}
and are the consequence of the lack of rotational symmetry in the Winfree and QIF models.

\begin{figure}[t]
\centerline{\includegraphics[width=85mm,clip=true]{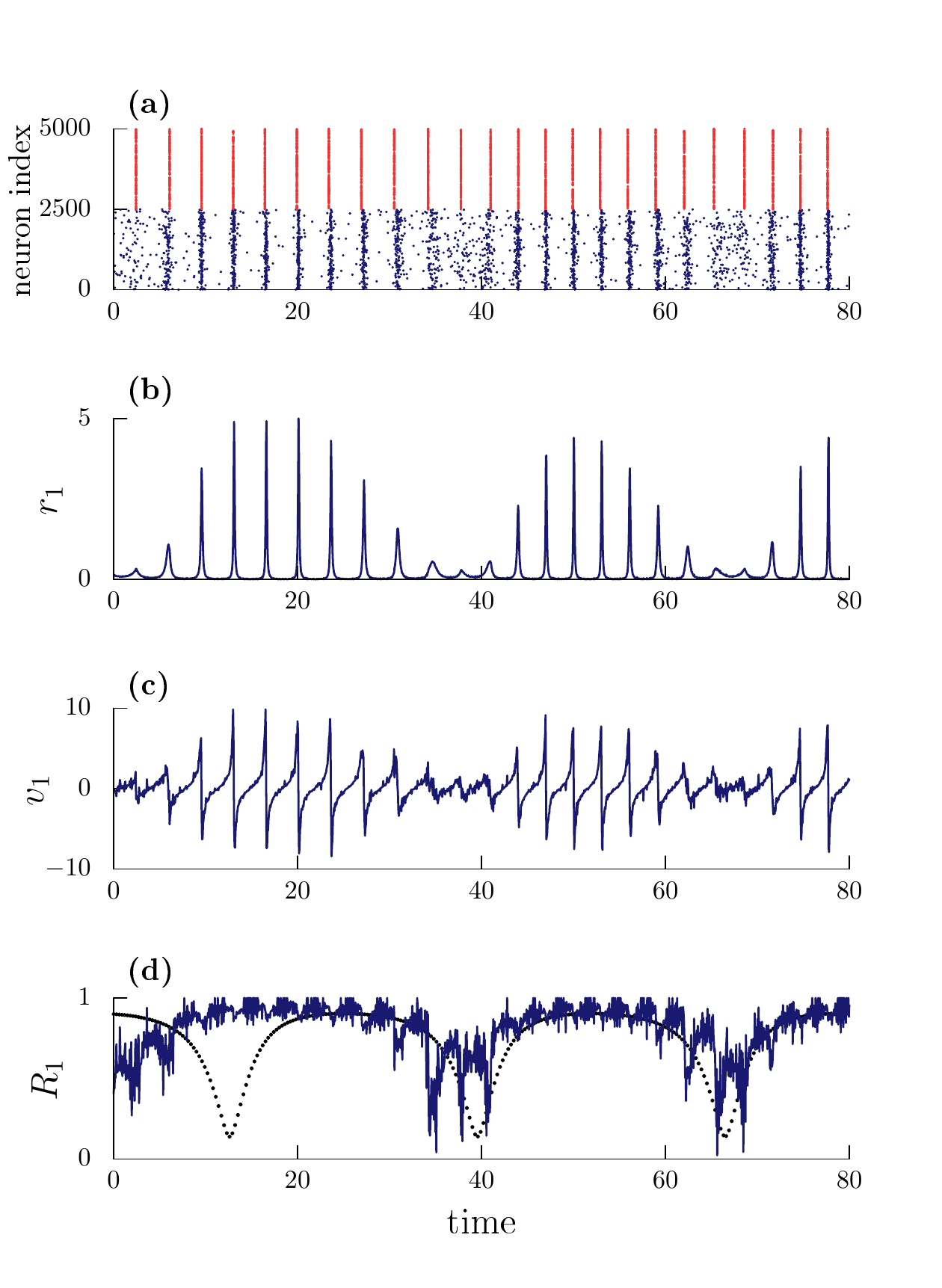}}
\caption{Quasiperiodic chimera state in a two-population network of $N=5000$ identical
inhibitory QIF neurons. The description of the panels and parameters is as 
in Fig.~\ref{Fig5} except $J_c/J_s=0.5$
(see also the $\times$ symbol in Fig.\ref{Fig3}b). 
}
\label{Fig6}
\end{figure}

In Fig.~\ref{Fig6}, we also explored how the unsteady chimera states 
in the Kuramoto model Eqs.~\eqref{KM2}  
translate to networks of QIF oscillators. 
To this aim we set the parameters of the QIF model
in the hatched region of the bifurcation diagram in Fig.~\ref{Fig3}b ($\times$ symbol), 
and used the same initial conditions as in the previous simulation.
Here, we find a more complex chimera state that seems to display  
macroscopic quasiperiodic dynamics
\footnote{We cannot discount that the collective dynamics 
of the QIF Eqs.\eqref{qif2} is chaotic, as it also occurs in coupled 
populations of Winfree oscillators~\cite{PM14}.}:
both the firing rate
and the mean-membrane potential of the incoherent population oscillate with 
two characteristic frequencies as can be appreciated in Figs.~\ref{Fig6}(b,c).
Again, the quasiperiodic chimera state in the QIF network corresponds to a 
periodic chimera state in the Kuramoto model. 
In Fig.~\ref{Fig5}(d) we used Eq.~\eqref{conformal} 
to represent the Kuramoto order parameter for the QIF network (blue), which 
roughly approximates the periodic dynamics of the corresponding 
Eqs.~\eqref{OA}.

\subsubsection{Chimeras for strong coupling}

The derivation of Eqs.~\eqref{KM2} from 
the QIF network Eqs.~\eqref{qif2} has been made under the assumption of 
weak coupling. Yet, are chimera states in QIF networks robust against 
stronger levels of coupling? 

To investigate this issue, we used the conformal map Eq.~\eqref{conformal} to express
the boundary of the basin of attraction of Fig.~\ref{Fig4}(a) in terms of
$(r_1,v_1)$. This transformed boundary is represented as a dashed line in Figs.~\ref{Fig4}(b-d).
Then, we performed three sets of numerical simulations of the QIF network for 
increasing values of $g_s$ while keeping the ratios $g_c/g_s$, $J_c/J_s$, and $\rho_s$ constant---note that this implies increasing all the other coupling parameters.

For very small values of $g_s$,
we expect the averaging approximation to hold, and so the stability boundary of 
chimera states in the QIF model. 
The green dotted region in Fig.~\ref{Fig4}(b) corresponds to the stability boundary 
of the steady chimera state in the QIF network
for $g_s=0.025$. The boundary approximately agrees with that of the KM (dashed line),    
although the region is slightly smaller in the QIF model.
Notably, further increases in coupling strength---see Figs.~\ref{Fig4}(c,d)---lead to a gradual reduction of the chimera's basin of attraction. 

To further investigate the reduction of the stability boundary of chimeras in the QIF model,
in Figs.~\ref{Fig4}(f-h) we computed the basins of attraction of chimera states 
in a different parameter regime---corresponding to the $+$ symbol in Fig.~\ref{Fig3}(a).
Here $|\rho_s|$ is three times larger than in the previous case, and hence,
chemical couplings are three times stronger than in Figs.~\ref{Fig4}(b-d). 
Correspondingly, the reduction of the basin of attraction of chimera states in 
Figs.~\ref{Fig4}(f-h) is more pronounced than in Figs.~\ref{Fig4}(b-d). 
In fact, in Fig~\ref{Fig4}(h), we find the complete disappearance of the region of stable 
chimera states in the QIF network. This suggests that chimera states in QIF networks are only 
observable for weak coupling. 
          
\section{Conclusions}

In this paper we have applied a perturbative approach to simplify
a weakly heterogeneous population of QIF neurons, with 
weak all-to-all chemical and electrical coupling Eqs.~\eqref{qif}. 
This approach leads to a classical variant of the Kuramoto model
Eq.~\eqref{KMp}~\cite{SSK88,SK86}, whose coupling parameters
satisfy a simple geometric relation with those of the QIF model
\footnote{A similar approach has been applied to the Leaky Integrate-and-Fire model. 
In this case the approximated phase model is not the KM, but it 
contains higher harmonics in the coupling function~\cite{PR15ii}}
Fig.~\ref{Fig1}.

The approximation of the QIF network by Eq.~\eqref{KMp} 
allows one to use the framework of the KM to investigate the 
role of chemical and electrical synapses in setting up synchronization. 
For example, we find that in the absence of electrical coupling  
the phase lag parameter of the KM is $\alpha=\pm \pi/2$, 
which prohibits synchronization; see also Ref.~\onlinecite{MP18}.  
Moreover, for Lorentzian distributions of currents 
the synchronization threshold depends only on electrical coupling, 
Eq.~\eqref{DeltaC}, whereas  
the oscillation frequency Eq.~\eqref{OmC} is determined by the ratio 
of chemical to electrical coupling. 
These results are in consonance with the exact description 
provided by so-called neural mass (or firing rate) models
for networks of QIF neurons~\cite{PDR+19,MP20}. 

The framework of the KM  
allows for uncovering and investigating dynamical states  
that are not reachable using neural mass models for QIF neurons~\cite{MPR15,Lai15,PDR+19,MP20}. 
Here, we analyzed the case of chimera states in two-population networks 
of identical QIF neurons~\cite{MKB04,AMS+08}. 
Despite the large number of studies devoted to investigate
chimera states in spiking neuron networks---see e.g.~\cite{OPT10,OOH+13,VHO+14,HKV+14,
PA15,MBG+19,Lai19,HKZ+16,MBG+19,GBS+20,LJH21}---,
the relation between such states and the original
chimera states uncovered in the KM~\cite{KB02,MKB04,AMS+08}
is lacking~\cite{BGL+20,Hau21}. We showed
that chimera states in QIF networks
emerge in the presence of both chemical and electrical couplings
but only if chemical coupling is much stronger than electrical coupling. 
However, our numerical results suggest
that chimeras in QIF networks are not robust against stronger levels of coupling. 

Finally, we introduced a framework for the analysis of QIF networks that can be 
readily applied to a variety of extensions of Eqs.~\eqref{qif}.
In particular, the derivation of Eq.~\eqref{KMp} does not 
impose constraints on the structure of the network or the 
shape of the distribution of heterogeneities. 
Given that network structure~\cite{SSK88,BT05}
and heterogeneities~\cite{Paz05,LCT10,OW12,MBS+09,PM09,PDD18} 
greatly affect the dynamics of the KM, 
it may be interesting to investigate how this translates to QIF networks.

\acknowledgments
The authors thank Diego Paz\'o for helpful discussions. 
PC acknowledges financial support from the European Union’s Horizon 2020 research and innovation programme under grant agreement No 101017716 (Neurotwin).
EM acknowledges support by the Agencia Estatal de Investigaci\'on 
under the Project No.~PID2019-109918GB-I00.

\section*{Data availability}

Data sharing is not applicable to this article as no new data were created or analyzed in this study.

\section*{Author Declarations}

The authors have no conflicts to disclose.

\bibliography{bibliografia}

\end{document}